\documentclass[10pt,letterpaper]{article}
\usepackage[top=0.85in,left=2.75in,footskip=0.75in,marginparwidth=2in]{geometry}

\usepackage[utf8]{inputenc}

\usepackage{ragged2e}

\usepackage{amsmath,amssymb,amsfonts}%
\usepackage{amsthm}%
\usepackage{mathrsfs}%

\usepackage{cite}

\usepackage{nameref,hyperref}

\usepackage[right]{lineno}

\usepackage{microtype}
\DisableLigatures[f]{encoding = *, family = * }

\raggedright
\usepackage{changepage}

\usepackage[aboveskip=1pt,labelfont=bf,labelsep=period,singlelinecheck=off]{caption}

\makeatletter
\renewcommand{\@biblabel}[1]{\quad#1.}
\makeatother

\usepackage{lastpage,fancyhdr,graphicx}
\usepackage{epstopdf}
\fancyheadoffset[L]{2.25in}
\fancyfootoffset[L]{2.25in}

\usepackage{color}

\definecolor{Gray}{gray}{.25}

\usepackage{graphicx}

\usepackage{sidecap}

\usepackage{wrapfig}
\usepackage[pscoord]{eso-pic}
\usepackage[fulladjust]{marginnote}
\reversemarginpar

\begin{document}
\vspace*{0.35in}

\begin{flushleft}
{\Large
\textbf\newline{Generation of protein conformations using a simple inverted transformer}
}
\newline
\\
Luigi Leonardo Palese\textsuperscript{1},
\\
\bigskip
\bf{1} Department of Translational Biomedicine and Neuroscience - University of Bari "Aldo Moro"
\\
\bigskip
* luigileonardo.palese@uniba.it

\end{flushleft}
\justifying
\section*{Abstract}
Molecular dynamics is an extremely powerful approach for studying protein behavior, but it requires enormous computational resources to obtain an adequate sampling of possible conformations. Recently, thanks to successes in predicting three-dimensional protein structures, there has been an explosion of research exploring the opportunities offered by deep learning to address the problem of protein conformational landscape. In this work, an artificial intelligence system based on a variant of the Transformer architecture is introduced and applied to the problem of determining the ensemble of three-dimensional structures that a protein can access and their statistical weight. After being trained on a dataset of conformations of a particular protein obtained using classical all-atom molecular dynamics, this neural network is able to directly generate physically plausible protein conformations similar to those obtained both through molecular dynamics and experimentally. The ordered sequence of conformations obtained as output of the neural network rapidly converges towards the most populated regions of the protein's conformational landscape, suggesting that the system may have learned not only the protein's three-dimensional organization but also the thermodynamic weights of the various feasible conformations. These features, combined with its simplicity and low computational load, make this Transformer variant an ideal candidate for the development of systems capable of performing generative molecular dynamics.
\section*{Introduction}
Proteins are molecules essential for cellular function, and knowledge of their structure and dynamics is the basis for understanding biological phenomena and improving biotechnological processes, including the development of new drugs. A thorough understanding of their function is therefore not only an academic topic, but also of practical interest. Protein dynamics is involved in the specific recognition of small molecules or macromolecular partners, enzymatic catalysis, and allosteric regulation \cite{hilser2012structural,henzler2007hierarchy,boehr2009role,henzler2007dynamic}. However, while the sequence and three-dimensional structure of a generic protein can be obtained relatively easily by the interested researcher by consulting the appropriate databases, determining its dynamics is an entirely different problem. Multiple conformations of proteins and, possibly, the time scales over which the conformational change occurs, can be obtained experimentally thanks to advances in acquisition and analysis of data from X-ray crystallography, nuclear magnetic resonance (NMR), and high-resolution electron microscopy, as well as more specialized techniques such as Förster resonance energy transfer \cite{fraser2011accessing,palmer2001nmr,hoff2024accurate,schuler2013single}. Alongside these experimental techniques, molecular dynamics (MD) \cite{karplus2005molecular,dror2012biomolecular,hollingsworth2018molecular} currently represents an important method for studying protein motion: in principle, the numerical resolution of the equations of motion obtained by MD provides both the conformational landscape of a protein and its dynamics \cite{kitao1999investigating,shao2007clustering,kitao2022principal,amadei1993essential,berendsen2000collective,atilgan2001anisotropy,bahar2005coarse}, in excellent agreement with experimental data. However, the experimental effort required for these analyses is considerable and the computational cost of massively sampling the entire conformational landscape of a protein still remains a formidable challenge, despite huge advances in both hardware and simulation algorithms \cite{stone2007accelerating}.

From these premises and inspired by the exceptional results in predicting static protein structures \cite{jumper2021highly,varadi2022alphafold,baek2021accurate,humphreys2021computed}, numerous computational approaches based on artificial intelligence (AI) algorithms have been developed to obtain information on protein dynamics at low computational cost. Despite being a very recent field of research, generative molecular dynamics (GenMD) already boasts numerous approaches \cite{dao2025deep,shi2026ai,janson2025generation,olsson2026generative}, ranging from simple language models to Boltzmann generators to diffusion models (see for example \cite{li2023exploring,noe2019boltzmann,zheng2024predicting,jin2025p2dflow,passaro2025boltz,lewis2025scalable,del2022sampling,wayment2024predicting,guan2024predicting}). The attention mechanism is capable of capturing the three-dimensional structure of proteins \cite{jumper2021highly}, as well as increasingly complex biophysical properties as the depth of the neural network increases, including binding sites and conformational ensembles of protein-protein complexes \cite{abramson2024accurate,wang2024exploring}. Transformer architectures \cite{vaswani2017attention} have proven to be effective in learning representations of protein structures useful for classification purposes, and networks of this type have also been used in the generative mode to predict protein dynamics and the underlying thermodynamic distribution, thus eliminating the need for classical MD  \cite{schwing2022molecular,zeng2021note,chennakesavalu2024data,jing2026ai,jing2024alphafold,lewis2025scalable}. Despite undoubted progress, current results are still far from allowing the prediction of the complete structural landscape of a protein (with the exception of a few toy models) and the thermodynamic weight of the various configurations. 

Questions remain regarding the actual ability to generate physically plausible sets of three-dimensional structures: an extremely important point, yet rarely highlighted in the GenMD literature, is that the output of the neural networks used in these AI-based algorithms is not a clean protein structure. Obtaining the final structure generally requires the use of algorithms that generate plausible protein conformations from the raw output of the neural network (see, for example, \cite{webb2016comparative} in \cite{schwing2022molecular} or \cite{singhal2025general} in \cite{lewis2025scalable}). Although the use of these algorithms is perfectly feasible in order to avoid meaningless structures as the final result, this suggests that the classical Transformer-based architectures currently in use are generally poorly suited to \textit{autonomously} predict a temporally ordered set of protein conformations. Moreover, the prediction power of the classical Transformer architecture in time forecast tasks has already been questioned \cite{zeng2023transformers} and it has been proven that even extremely complex architectures are not capable of providing performance superior to simple linear layers, similar to the statistical forecasters introduced a long time ago \cite{box1968some}.

In this work, it is demonstrated that a variant of Transformer can be rapidly trained to generate three-dimensional coarse-grained protein structures without having to resort to systems downstream of the neural network output. This neural network directly produces ordered sequences of protein conformations that are statistically distributed in a manner similar to those obtained by MD. This expands the number of available conformations at a very low computational cost and could also be a key factor in the development of future AI systems capable of predicting protein dynamics without resorting to MD itself. Dihydrofolate reductase (DHFR) was chosen as the model system because it is one of the best-characterized protein due to its central role in metabolism \cite{schnell2004structure}. It is an ubiquitous enzyme that catalyses the reduction of 7,8-dihydrofolate (DHF) to 5,6,7,8-tetrahydrofolate (THF) via a proton-coupled hydride transfer, using nicotinamide adenine dinucleotide phosphate (NADPH) as a cofactor. Its role in maintaining the intracellular pool of THF, which in turn is essential for the synthesis of purines and thymidylate, has justified numerous studies on this enzyme as a target for anticancer drugs and antibiotics \cite{schweitzer1990dihydrofolate,hawser2006dihydrofolate,raimondi2019dhfr}. Moreover, DHFR is a paradigmatic example of the role of dynamics in enzymatic catalysis fine-tuned for specific cellular environments \cite{schnell2004structure,verma1997domain,sham2001molecular,cetin2023kinetic,bhabha2013divergent}, and the availability of experimental data makes this enzyme very attractive for MD studies, including the development of new MD analysis algorithms \cite{love2023evaluating,kalmer2024statistical}. 

\section*{Methods}
\subsection*{Molecular dynamics}
The atomic coordinates of \textit{E. coli} DHFR were obtained from PDB~\cite{berman2000protein,burley2021rcsb} entry 3QL3, which also contains as ligands nicotinamide adenine dinucleotide phosphate and folic acid~\cite{bhabha2011dynamic}. Structure parameterization was carried out using CHARMM-GUI~\cite{jo2008charmm,brooks2009charmm,lee2016charmm}. The structure of apoprotein was obtained by removing the ligands from the binding site, while the starting structures of the protein bound to NADP alone or to NADP and folic acid were obtained by keeping only NADP or both ligands in the 3QL3 structure. Ligand parameterization was performed using the CHARMM General Force Field to generate the topology and parameter files starting from the coordinates reported in the PDB \cite{kim2017charmm}. These systems were immersed in a water box with 20~\AA~ of padding containing 0.15M KCl under charge-neutralizing conditions; the pH of the system was set at 7.0.

MD was performed in NAMD \cite{phillips2020scalable} (version 3.0 compiled using GCC version 12.2.0 and CUDA 12.2) on an HPC system, using the CHARMM36m force field~\cite{huang2017charmm36m}, periodic boundary conditions and particle-mesh Ewald (PME). The~TIP3P water model~\cite{jorgensen1983comparison} was used. Pairs of bonded atoms were excluded from non-bonded interactions by setting the \textit{exclude} parameter to \textit{scaled1-4} in the NAMD configuration file, and van der Waals interactions were truncated at 12~\AA~ using the built-in NAMD function that smoothly truncates the van der Waals potential energy at the cutoff distance (\textit{switchdist} parameter 10.0). Electrostatics and van der Waals interactions were calculated for all pairs of atoms whose distance was less than or equal to 16~\AA, and reassignment of the atom lists for the calculation of non-covalent interactions was performed every 10 time steps. Short range non-bonded interactions and long-range electrostatic interaction forces were evaluated at each time step. After 10,000 conjugate gradient minimization steps, systems were equilibrated in the NVT ensemble for 5 ns with a 2 fs time step at 300~K using Langevin dynamics; a Langevin coupling coefficient equal to 1/ps was applied to all atoms, but excluding hydrogens. After equilibration, the production run was performed for 100 ns under NPT conditions. Langevin piston Nose-Hoover method in NAMD was used for pressure control \cite{martyna1994constant,feller1995constant} (target pressure 1.01325 bar, barostat oscillation time scale 100~fs, barostat damping time scale 50~fs).

\subsection*{Computational methods}
Computational scripts for molecular trajectory extraction and prediction were generated and iteratively refined through interaction with the large language model Gemini (Google). All AI-generated outputs, including the attention-to-atom reduction logic and the recursive sliding-window protocol, were manually audited, debugged for physical consistency (e.g., C$\alpha$-C$\alpha$ bond distances), and validated against ground-truth MD simulations by the author.

\subsubsection*{Preparing the dataset}
For the \textit{E. coli} DHFR apoprotein, two independent datasets were obtained as reported above, each containing 30 MD simulation runs, each lasting 100 ns.
For the following analyses and for displaying the results the MDAnalysis, VMD, Numpy and Matplotlib software libraries were used \cite{HUMP96,gowers2019mdanalysis,michaud2011mdanalysis,harris2020array,hunter2007matplotlib}. Each of the simulations was aligned to the same reference frame to eliminate global rotational and translational motions of the protein. The trajectories were initially aligned using the fast QCP algorithm to compute the root mean square distance (RMSD) between two sets of coordinates \cite{theobald2005rapid} and the rotation matrix $\mathsf{R}$ and the translation vector $\mathbf{t}$ that minimizes the RMSD were calculated \cite{liu2010fast}. The Cartesian coordinates of the C$\alpha$ atoms were extracted, arranged in a matrix of size $N_{frame} \times (atoms*3)$ and saved in Numpy's NPZ format, but excluding the equilibration frames. Subsequently, the RMSD was calculated between the coordinate matrices of each time sample $x$ and $x_{ref}$ according to the following equation
\begin{equation}
    \text{RMSD}(\mathbf{x}, \mathbf{x}^{\text{ref}}) = \min_{\mathsf{R}, \mathbf{t}} %
 \sqrt{\frac{1}{N} \sum_{i=1}^{N} \left[ %
     (\mathsf{R}\cdot\mathbf{x}_{i}(t) + \mathbf{t}) - \mathbf{x}_{i}^{\text{ref}} \right]^{2}}
\end{equation}
The RMSD of the C$\alpha$ atoms was calculated using the initial simulation frame as reference. The root mean square fluctuation (RMSF) of a structure is the time average of the RMSD. The RMSF was calculated using the following equation
\begin{equation}
    \rho_i = \sqrt{\left\langle (\mathbf{x}_i - \langle\mathbf{x}_i\rangle)^2 \right\rangle}
\end{equation}
where $x_i$ are the coordinates of particle $i$ and $\langle x_i \rangle$ is the ensemble-averaged position of $i$.
The dynamic cross-correlation matrix (DCCM) \cite{swaminathan1991investigation} was calculated on each production run using the following equation:
\begin{equation}
C_{ij} = \frac{\sum_{t=1}^{T} (\mathbf{r}_i(t) - \langle \mathbf{r}_i \rangle) \cdot (\mathbf{r}_j(t) - \langle \mathbf{r}_j \rangle)}{\sqrt{\left[ \sum_{t=1}^{T} (\mathbf{r}_i(t) - \langle \mathbf{r}_i \rangle)^2 \right] \left[ \sum_{t=1}^{T} (\mathbf{r}_j(t) - \langle \mathbf{r}_j \rangle)^2 \right]}}
\end{equation}
where $\Delta \mathbf{r}_i(t) = \mathbf{r}_i(t) - \langle \mathbf{r}_i \rangle$ represents the fluctuation vector for each time instant $t$ with respect to the average position calculated over the entire trajectory. Principal component analysis (PCA) was performed by solving the eigenvalue problem of the covariance matrix whose elements $(i, j)$ are the covariance between the coordinates $i$ and $j$. All these analyses were performed using the appropriate built-in functions in MDAnalysis \cite{gowers2019mdanalysis,michaud2011mdanalysis}.

\subsubsection*{Model architecture: \texttt{iTransformer} for molecular dynamics}
To model long-range protein trajectories and capture the temporal evolution of atomic coordinates, an adapted version of the inverted Transformer architecture \cite{liu2023itransformer} has been implemented, which addresses the limitations of classical Transformers in multivariate time series forecasting. This architecture will be called \texttt{iTransformer}, hereafter referred to as \texttt{iT}. Unlike traditional time-series models, the \texttt{iT} processes entire multivariate time series by treating each structural variable (atomic coordinates) as an independent token. 

Let the input sequence be represented as a sliding window of normalized spatial coordinates $X \in \mathbb{R}^{L \times V}$, where $L$ denotes the \texttt{lookback} history length and $V$ represents the number of variates (corresponding to $3 \times N_{\text{atoms}}$). The input tensor undergoes a permutation layer to invert time and variable axes:
\begin{equation}
    X_{\text{perm}} = \text{Permute}(X) \in \mathbb{R}^{V \times L}
\end{equation}
A dense embedding layer projects the inverted representation into a latent space of dimension $d_{\text{model}} = 128$:
\begin{equation}
    H_0 = \text{Dense}(X_{\text{perm}}) \in \mathbb{R}^{V \times d_{\text{model}}}
\end{equation}
The core architecture consists of specialized \texttt{iTransformerBlock} layers featuring multi-head self-attention mechanisms (in all the results shown in this work the parameter $num\_heads = 8$ was used) combined with layer normalization and position-wise feed-forward networks (FFN with a GELU activation function \cite{hendrycks2016gaussian} and hidden dimension $ff\_dim = 512$). This mechanism captures global inter-variable correlations (cooperative motions among atoms):
\begin{equation}
\mathbf{A} = \text{Attention}(\mathbf{Q}, \mathbf{K}, \mathbf{V}) = \text{softmax}\left(\frac{(\mathbf{H}\mathbf{W}^Q)(\mathbf{H}\mathbf{W}^K)^\top}{\sqrt{d_k}}\right)(\mathbf{H}\mathbf{W}^V)
\end{equation}
where $\mathbf{H} \in \mathbb{R}^{V \times d_{model}}$ is the matrix of variate embeddings and $\mathbf{A} \in \mathbb{R}^{V \times V}$ is the attention matrix (technically $d_k$ is related to the size of the matrix $K$ and here it is given by the size of the variates). The attention weights effectively represent the dynamic coupling between different residues, capturing both short-range covalent constraints and long-range non-bonded interactions inherent in the protein fold. This variate-wise attention provides an interpretable map of how the motion of a specific residue is conditioned by the rest of the structure.
Finally, a temporal projection layer maps the features back to the required forecast horizon $F$, followed by a reverse permutation to yield the predicted trajectory block $\hat{Y} \in \mathbb{R}^{F \times V}$.

During the tests carried out to optimize the neural network, it was seen that one of the most important parameters is the dropout \cite{srivastava2014dropout}. Dropout is a regularization technique in neural networks that randomly sets a fraction of the neurons' output to zero during each training phase. This technique reduces complex co-adaptations between neurons, forcing the network to learn more robust and generalized features, thus preventing overfitting. In particular, this parameter forces the Transformer to ignore background noise due to random phenomena and to try to focus its attention on parameters that allow it to generalize better. In this work, dropout values ranged from 0.1 to 0.3 (with most reported data referring to systems trained with dropout=0.1 unless specifically stated otherwise).

\subsubsection*{Custom RMSF-weighted hybrid loss function}
To prevent the neural network from smoothing out physical fluctuations over long prediction horizons—a common issue in standard Mean Squared Error (MSE) optimizations—a custom hybrid loss function was introduced. This loss combines standard pointwise accuracy with a structural fluctuation penalty that mimics the Root Mean Square Fluctuation (RMSF).

Given the ground truth trajectory batch $Y_{\text{true}}$ and the predicted trajectory $\hat{Y}$ of shape $(B, F, V)$ where $B = \text{batch\_size}$, the hybrid loss is defined as:
\begin{equation}
    \mathcal{L}_{\text{total}} = \mathcal{L}_{\text{MSE}} + \alpha \mathcal{L}_{\text{RMSF}}
\end{equation}
where $\mathcal{L}_{\text{MSE}}$ is the standard mean squared error computed pointwise:
\begin{equation}
    \mathcal{L}_{\text{MSE}} = \frac{1}{B \cdot F \cdot V} \sum_{b, f, v} (Y_{\text{true}, b,f,v} - \hat{Y}_{b,f,v})^2
\end{equation}
and $\mathcal{L}_{\text{RMSF}}$ penalizes discrepancies in temporal fluctuations evaluated along the forecast horizon axis (axis $1$):
\begin{equation}
    \text{RMSF}_{\text{true}} = \sigma_t(Y_{\text{true}}) = \sqrt{\frac{1}{F}\sum_{f=1}^{F} (Y_{\text{true}, f} - \mu_t(Y_{\text{true}}))^2}
\end{equation}
\begin{equation}
    \text{RMSF}_{\text{pred}} = \sigma_t(\hat{Y}) = \sqrt{\frac{1}{F}\sum_{f=1}^{F} (\hat{Y}_f - \mu_t(\hat{Y}))^2}
\end{equation}
\begin{equation}
    \mathcal{L}_{\text{RMSF}} = \frac{1}{B \cdot V} \sum_{b, v} (\text{RMSF}_{\text{true}, b,v} - \text{RMSF}_{\text{pred}, b,v})^2
\end{equation}
The hyperparameter $\alpha$ balances the trade-off between absolute coordinate accuracy and the preservation of conformational fluctuation dynamics.

\subsubsection*{Curriculum learning strategy}
Training autoregressive models directly over a long forecast horizon often suffers from error accumulation and gradient instability. To ensure robust convergence, it has been implemented a curriculum learning strategy \cite{bengio2009curriculum}. 

The training process is divided into sequential stages with progressively increasing forecast horizons (the \texttt{forecast} parameter), starting from an initial horizon (e.g., $10$ steps) up to the final target horizon (e.g., $50$ or more steps) with $f_{\text{steps}}$ incremental steps. For each stage, the model is initialized with the optimized weights of the previous stage (Warm Start). Optimization is performed using the AdamW optimizer \cite{loshchilov2017decoupled} coupled with a Cosine Decay learning rate scheduler (initial learning rate $10^{-4}$, weight decay $10^{-3}$) across multiple epochs per stage, ensuring stable adaptation from short-term dynamics to long-range trajectory rollouts.

\subsubsection*{Temporal projection and multistep forecasting}
Following the attention and feed-forward refinement, the refined latent representations are mapped back to the physical coordinate space. Instead of an iterative autoregressive decoding, it is used a direct \textit{temporal projection}. Each latent token $\mathbf{z}_v$ is independently projected to the forecast horizon $F$ (in this work $10 \leq F \leq 100$) using a linear layer:
\begin{equation}
\hat{y}_{1:F, v} = \mathbf{W}_{proj} \mathbf{z}_v + \mathbf{b}_{proj}
\end{equation}
This projection interprets the encoded "state" of each atom to predict its future trajectory segment. The final output is permuted back to $\hat{\mathbf{Y}} \in \mathbb{R}^{F \times V}$, yielding a continuous $F$-frame prediction of the protein's conformational evolution. This direct mapping significantly reduces cumulative error compared to step-by-step forecasting and ensures computational efficiency during large-scale inference tasks.

\subsubsection*{Attention matrix visualization}
The following procedure was used to obtain the attention matrix images. The \texttt{iT} was initialized with previously calculated weights and used in generative mode starting from a given \texttt{loockback} sequence. The attention matrix weights were then extracted. From the raw attention matrix of dimension $(num\_heads, V, V)$ the average with dimension $(V, V)$ was calculated. 
To move from a representation of degrees of freedom ($V$) to the number of atoms ($N_{\text{atoms}}$), a four-dimensional reshape was performed obtaining a tensor $(N_{\text{atoms}}, 3, N_{\text{atoms}}, 3)$ and then the average was calculated along the $x$, $y$, and $z$ spatial dimensions, corresponding to axes 1 and 3 of the previous tensor in python-style numbering. The result is a final $N_{\text{atoms}} \times N_{\text{atoms}}$ square matrix, where each cell represents the combined average attention intensity between the source and target atoms. The attention matrix was saved sequentially for each desired frame interval and displayed as a color map.

\subsubsection*{The \texttt{iT} model in generative mode}
To calculate the pseudo-trajectory predicted by an \texttt{iT} model, a function capable of recursively generating protein conformations in an autoregressive manner was implemented. Since the \texttt{iT} model is trained to predict a fixed time horizon (defined by the \texttt{forecast} parameter, set from 10 to 100 in the tests shown in this work), to generate longer trajectories (up to 1000 steps) the function proceeds iteratively: at each cycle it uses the newly generated predictions as input to estimate the subsequent steps (sliding window). The function takes the following arguments: the already trained model, the initial historical time window having size $(L, V)$ where $L$ is the lookback and $V$ is the number of variables ($N_{\text{atoms}} \times 3$), and the number of future frames to be recursively generated. At this point, the cycle begins and will be repeated recursively: a batch size (needed by Keras) is added to the current window (which goes from $(L, V)$ to $(1, L, V)$); this is fed as input to the model, which outputs the next time block. The raw output of the \texttt{iT} has size $(1, F, V)$ where $F$ is the forecast parameter; the batch size is removed and the block of size $(F, V)$ is saved in the list of generated frames. At this point, the autoregressive sliding window technique is used: the old time window and the new predicted block are concatenated along the time axis (resulting in a matrix of size $((L + F), V)$). This matrix is sliced to obtain a new window of size $(L, V)$ containing the last $L$ elements of the output. The cycle ends when the number of generated conformations is equal to or greater than the requested number. All the chunks accumulated in the list of generated frames are concatenated along the time axis, obtaining a matrix of size $(\textit{number of cycles} \times F, V)$, possibly truncated to the total number of steps required (final returned dimension: $(T, V)$ where $T$ is the number of total steps and $V$ is the number of variables).

This matrix, which contains the Cartesian coordinates of the selected atoms (the C$\alpha$ atoms in the cases presented in this communication), was used to visualize the pseudo-dynamics of the protein in the space defined by the first two PCs of the simulation from which the initial time window was obtained (through a simple projection into this space). The same matrix was used to obtain the Cartesian coordinates of the C$\alpha$ atoms to be inserted into a \textit{pdb} file, in order to visualize the spatial arrangement of the selected atoms in the conformations predicted by the \texttt{iT}.

The probability distribution function of the conformations obtained by MD or by recursive use of \texttt{iT} in a generative way was calculated via the kernel density estimation function in Scipy \cite{Virtanen2020}.

\section*{Results}
\subsection*{Molecular dynamics and dataset construction}
To build the dataset, the \textit{Escherichia coli} DHFR structure was used: 66 different simulations were performed, each of which, after the minimization and equilibration steps, yielded a 100 ns simulation sampled at 10 ps. These simulations were arbitrarily divided into two groups (from here onward, \textit{dataset1} and \textit{dataset2}), each containing 30 simulations, leaving out six simulations for further analysis (never seen by the implemented AI systems). Each of these two groups was divided into a training set containing 25 simulations, a validation set containing 4 simulations, and a test simulation; these groups were kept separate for all analyses. For all these simulations, the RMSD, RMSF, DCCM, and PCA were calculated (see Supplementary Figures 1 and 2 in \textit{Supporting Information} where two examples of this type of analysis are reported). Alongside these two large datasets, several smaller datasets were constructed to test the AI system's ability to learn protein dynamics using a smaller amount of data (e.g. \textit{small-dataset1}). For further details on the composition of the datasets used, the interested reader may consult the \textit{Supporting Information}.

Inspection of the RMSD data suggests that the protein visits multiple conformations and maintains them for more or less long periods of time. This is confirmed by the PCA plots: while some simulations seem to explore a single minimum, for the majority of them at least two (and up to four) different regions appear to be explored. This is perfectly in line with what is expected for this enzyme, for which a catalytic cycle containing several intermediates, some of which are structurally distinct, has been described. Although these intermediates refer to the enzyme bound to substrates or products (or inhibitors), conformational flexibility is expected even in the absence of ligands \cite{boehr2006dynamic,hammes2009conformational,vogt2013conformational,weikl2014conformational}. These characteristics make the DHFR an ideal model for evaluating the possibility of training an AI system on protein dynamics: it is stable enough to present a recognizable three-dimensional structure but at the same time not so static and rigid as to induce the neural network to respond in a trivial way (i.e. by exactly repeating the last structure learned or an average of the structures in the training set, with or without the addition of noise). The similarity of the DCCM patterns and the RMSF profiles obtained in the various simulations suggests that the complex protein movements are based on reciprocal interactions between residues, interactions that can even be distant in space and time. Although this consideration may seem trivial in the context of protein dynamics, it is fundamental to the purpose of this work since the goal is to verify the feasibility of the analysis and prediction of complex systems characterized by non-linear and time-varying (but not random) relationships using an AI model.

\subsection*{The inverted Transformer}
Once it was established that the target protein used as a model was suitable for this type of research and that the set of simulations visits an adequate number of conformations and transitions between conformational basins, a series of neural networks have been implemented with the aim of obtaining a lightweight and easily trainable AI system, capable of directly providing as output an ordered series of physically plausible conformations of the protein of interest. Transformer architectures \cite{vaswani2017attention}, often coupled with rotationally invariant representations (based, for example, on the $\phi$ and $\psi$ angles) have proven to be effective in learning representations of protein structures in state-of-the-art systems (see Introduction section). In the context of protein dynamics neural networks of this type appear to be trainable, with (at least apparently) satisfactory error metrics and quite plausible Ramachandran plots (see, for example, \cite{schwing2022molecular}). But the raw output of the Transformer module consists of completely non-physical protein structures, with a variety of gaps, overlaps, and steric hindrances, and even distortions of the secondary structure motif, requiring the intervention of specialized software downstream of the Transformer as mentioned in the Introduction. This suggests that the classical Transformer architecture is not suitable for implementing an AI system that is lightweight enough to be trained without resorting to HPC systems but still capable of providing physically acceptable structures as direct output. Thence the idea of changing the perspective of the tokens used by the neural network, breaking away from the typical Transformer implementation.

The Transformer architecture is intimately linked to the need to transform words in linguistic models into a vector (or, more generally, a tensor). In the classical setup for time series forecasting the Transformer takes as input time tokens which are nothing more than a \textit{multivariate} representation of a single time step \cite{wu2021autoformer,zhou2021informer}. To prevent the vocabulary from becoming infinite, coarse-grained representations of the system are often introduced, resulting in a pseudo-vocabulary so that the system dynamics can be treated as a linguistic system \cite{schwing2022molecular}. The proposed change of perspective was to use \textit{univariate} time series as tokens, instead of multivariate time steps. In practice, within the context of MD, this translates into using the time-series vectors of the various degrees of freedom characterizing the protein's motion as tokens. Channel independence has been proposed in some attention-based neural network architectures, but introducing significant modifications to the Transformer module (see, for example, \cite{nie2022time}). However, in the Liu \textit{et al.} approach \cite{liu2023itransformer}, the Transformer architecture is kept essentially intact. Here, a similar inverted Transformer was implemented, using the individual degrees of freedom of the protein (i.e., the Cartesian coordinates $x$, $y$, and $z$ for each atom considered in the dataset) as tokens. The entire time series for each degree of freedom, in each of the simulations contained in a specific dataset, was divided into fragments containing a certain number of past time steps (the \texttt{lookback} sequence) that the algorithm uses to predict a certain number of future time steps (the \texttt{forecast} sequence). The algorithm implemented in this work is particularly simple as it does not even consider positional embedding or amino acid coding systems, which are typically included in neural networks containing Transformer modules reported in the protein literature. Although this (intentionally) light architecture as such does not allow for the system's generalization (i.e. it is not possible to train such a system with one protein and then apply it directly to a different one), this does not exclude that such an algorithm could be extended by inserting positional and amino acid type encoding, but this was outside the scope of this communication. A detailed description of the algorithms implemented in this work is reported in the Methods section, as well as the code needed to reproduce the obtained data. The following refers to a Transformer architecture with 8 heads, input embedding dimension of 128, feed-forward dense layer of dimension 512 and dropout 0.1 or 0.3.

\subsection*{How an inverted Transformer sees protein dynamics}
After training an \texttt{iT} network on \textit{E. coli} DHFR MD datasets, attention matrix images were obtained that describe which amino acid residues the neural network focuses on to predict the future evolution of protein dynamics. Figure 1 shows two such images: they are the attention matrices obtained by providing as input a particular temporally ordered sequence of conformations taken from an MD simulation (the \texttt{lookback} sequence). 
To avoid any possible unintentional data leakage, the attention matrices shown in the Figure were obtained using one of the simulations of the group never seen by the Transformer (i.e. not part of the training, validation, or test sets). Considering the structural organization of DHFR and its dynamics, one (naively) expected an attention matrix organized in blocks and diagonals, reflecting the presence of two subdomains, namely the adenosine-binding subdomain and the major subdomain, and of functionally important regions such as the Met20 loop, and the hinge residues Lys38 and Val88. Residues belonging to one subdomain could look to residues in the same subdomain for accurate prediction of the same subdomain motions, and long-range correlations could plausibly be expected for residues such as those belonging to loop or hinge regions. Diagonals were also expected because it would seem intuitive to think that, to predict the dynamics of a particular residue $i$, the system would focus its attention on the residue $i$ itself and its neighboring ones as well as residues distant in the sequence but close in the three-dimensional structure, similarly to what can be seen in correlation matrices. But attention matrix organization of \texttt{iT} looks completely different from the correlation matrices: the system's attention focuses on very few residues, and these residues guide the prediction of the dynamics of practically all the others. Hence, the presence of a vertical line around a particular residue $i$ with values that are not exactly equal for each residue $j$, but still on average higher than the attention reserved by $j$ for a generic residue. Analysing the entire series of time frames of the simulation with a particular \texttt{iT}, it can be observed (not shown) that transiently the attention is  directed to other residues (weakly in some cases), but also in this case all the residues of the protein pay attention to some particular residues for the prediction of the whole dynamics, preserving this aspect of the attention matrix dominated by a few vertical lines. As can be seen from the Figure, despite using datasets of different sizes and a different training scheme with a different number of allowed epochs, the appearance of the attention matrix obtained with the two iTs is very similar, apart from a little more noise in the matrix obtained using the \textit{mini-dataset1}.

\begin{figure}[ht] 
\includegraphics[width=\textwidth]{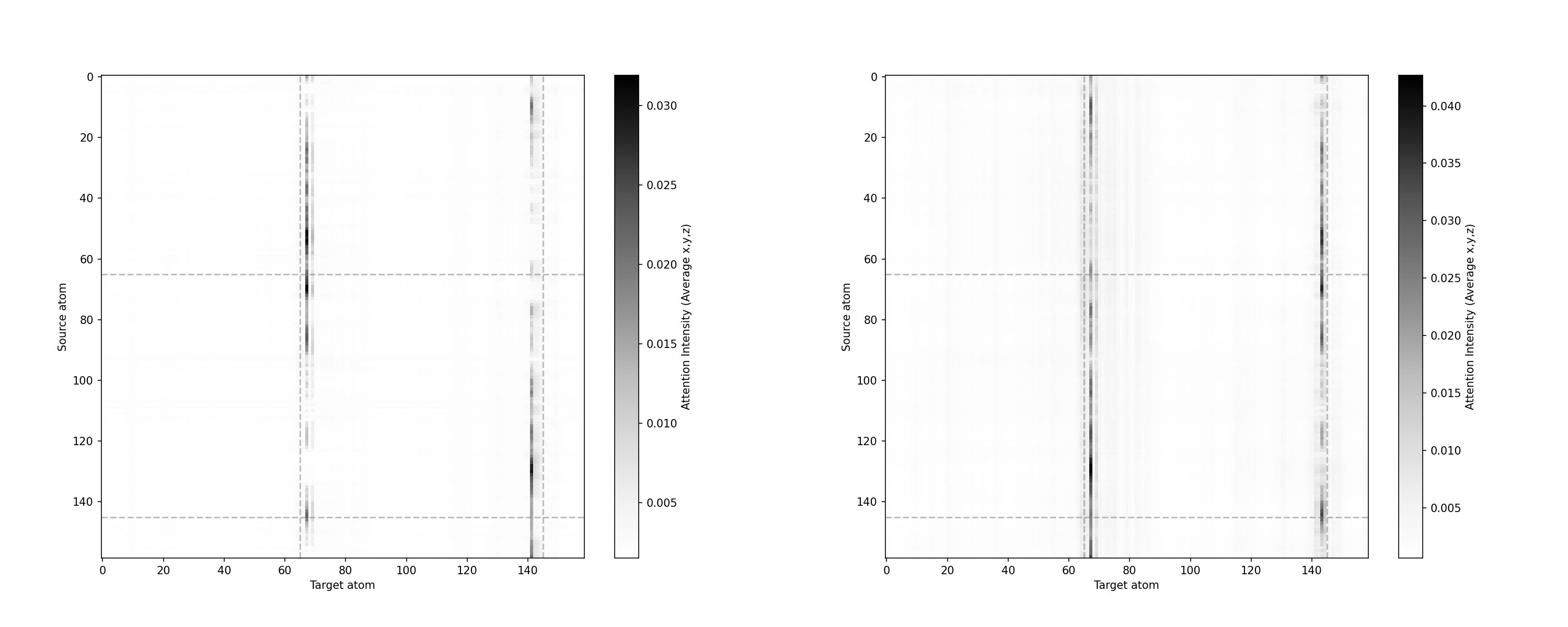}
\caption{\color{Gray} \textbf{The attention matrix.} The attention matrices shown in the figure were both obtained using the sequence of frames from run \#61 ending with frame 3500 as \texttt{lookback}. The attention matrix on the left was obtained using the weights of an \texttt{iT} trained on \textit{dataset1} (containing 30 elements; MAE 0.6710). The attention matrix on the right was obtained by training the \texttt{iT} on a \textit{small-dataset1} (containing 7 elements; MAE 0.6744). In both cases, training was performed using the custom loss function and the learning curriculum until a \texttt{lookback}/\texttt{forecast} ratio of 60/100 was obtained (see Methods section). The maximum number of epochs allowed for each training phase was 50 for the \texttt{iT} trained on the larger dataset and 10 for the \texttt{iT} trained using the small dataset. The dashed lines are used only as an aid to assess the attention paid from all residues $j$ of the protein sequence to residue $i$ (vertical lines) or from residue $i$ to all residues $j$ (horizontal lines) and are centered on residues 65 and 145.}
\end{figure}

The appearance of a few vertical lines in the attention matrix, where the \texttt{iT} focuses to predict the dynamics of the entire protein, was an unexpected and puzzling finding of this work. The residues on which attention appears to be focused are not particularly conserved in evolution, and do not appear to be associated with mutations that alter the kinetics or catalytic mechanism or the structural stability of the enzyme. One possible interpretation of the result could be that the high-attention residues are high-mobility ones and therefore this is the result of the neural network's attempt to focus on regions with greater conformational variance to reduce the prediction error, considering that more rigid portions of the protein are relatively constant and repeating their structure or simply averaging it may be sufficient to achieve good predictive performance. To evaluate this hypothesis, the RMSF of the high-attention regions was analysed. Indeed, the high-attention residues are in peripheral loops that generally show a high RMSF in the \textit{E. coli} protein, but upon further analysis, this simple explanation does not hold. First, these residues are not exactly the RMSF maxima, as they should be if the above hypotheses were correct. Furthermore, there are regions with an RMSF equal to or even higher than that on which the Transformer's attention is focused, but which are characterized by very low attention. For example, the DHFR region around residues 18-20 generally shows an RMSF equal to or higher than the region around residue 65 (see Supplementary Figures 1 and 2 in \textit{Supporting Information}), but it is never a region of high attention (with all iTs tested regardless of the training dataset used; not shown). In other words, some high mobility loops may be important for predicting the dynamics of the entire protein, but not all loops are equally important. Furthermore, attention is not focused on the entire loop but only on some residues. Moreover, the same attention pattern (few vertical lines in which the largest part of attention is concentrated) is also observed in other proteins that have been tested (human DHFR and phage T4 lysozyme; data not shown). After excluding a computational artifact like the "attention sink" of the Transformer by performing specific tests of anomalous accumulation of attention on spurious degrees of freedom (not shown),  the most likely explanation for the appearance of the attention matrices is that the \texttt{iT} model, not being natively equivariant, must learn by itself and construct an internal coordinate system using two or three residues as fixed anchors on which to superimpose and align all the chain configurations.

\subsection*{Generated conformational ensembles}
In the experiments illustrated in the previous section, the \texttt{iT} is used analytically, providing a series of \texttt{lookback} sequences from a MD run to obtain the corresponding attention matrices. The trained \texttt{iT} can also be used in generative mode. In this type of experiment, a sequence of protein conformations obtained by MD is used as a primer sequence (\texttt{lookback} sequence) that leads to the prediction of a certain number of frames. In turn, the predicted frames are used as primers until the required number of conformations is produced (thus obtaining a pseudo-molecular dynamics, a sort of GenMD). To better evaluate the structures generated via \texttt{iT}, the predictions of the neural network in generative mode were compared with structures obtained through MD. To this end, the predicted structures were projected onto the principal components (PCs) calculated from the simulation that provided the \texttt{lookback} sequences. This analysis reveals that the \texttt{iT} in generative mode produces a sequence of structures that, starting from a given region of the protein's conformational landscape, tends to reach regions densely populated in the MD simulation, following a trajectory that unfolds along the first two PCs, often smoothly. The specific outcome obviously depends on the \texttt{iT} weights employed (and thus on the specific training performed), as well as on the seed sequences used for prediction. Figure 2 shows the results obtained using weights derived from the training performed on the two largest datasets. Using the same \texttt{lookback} sequence as a seed, the sequences of conformations predicted by the two iTs are very similar. It should be noted that the dynamics used for this experiment belong to the group of simulations not used to train the neural network (i.e., they are not included in either \textit{dataset1} or \textit{dataset2}). The MAE, calculated using the test simulations contained in the two datasets, was 0.6710 for \texttt{iT} trained on the \textit{dataset1} and 0.6902 for the \texttt{iT} trained on the \textit{dataset2}. Although the trajectories are not identical, in both cases the process begins in a peripheral region of the conformational landscape and ends in a region very close to the origin. In addition, the shape of the curve traced by the two GenMD systems is strikingly similar. It should be noted that the starting points of the two predicted trajectories are very similar but not identical: this is because the Figure displays the predictions generated by the two iTs.
\begin{figure}[ht] 

\includegraphics[width=\textwidth]{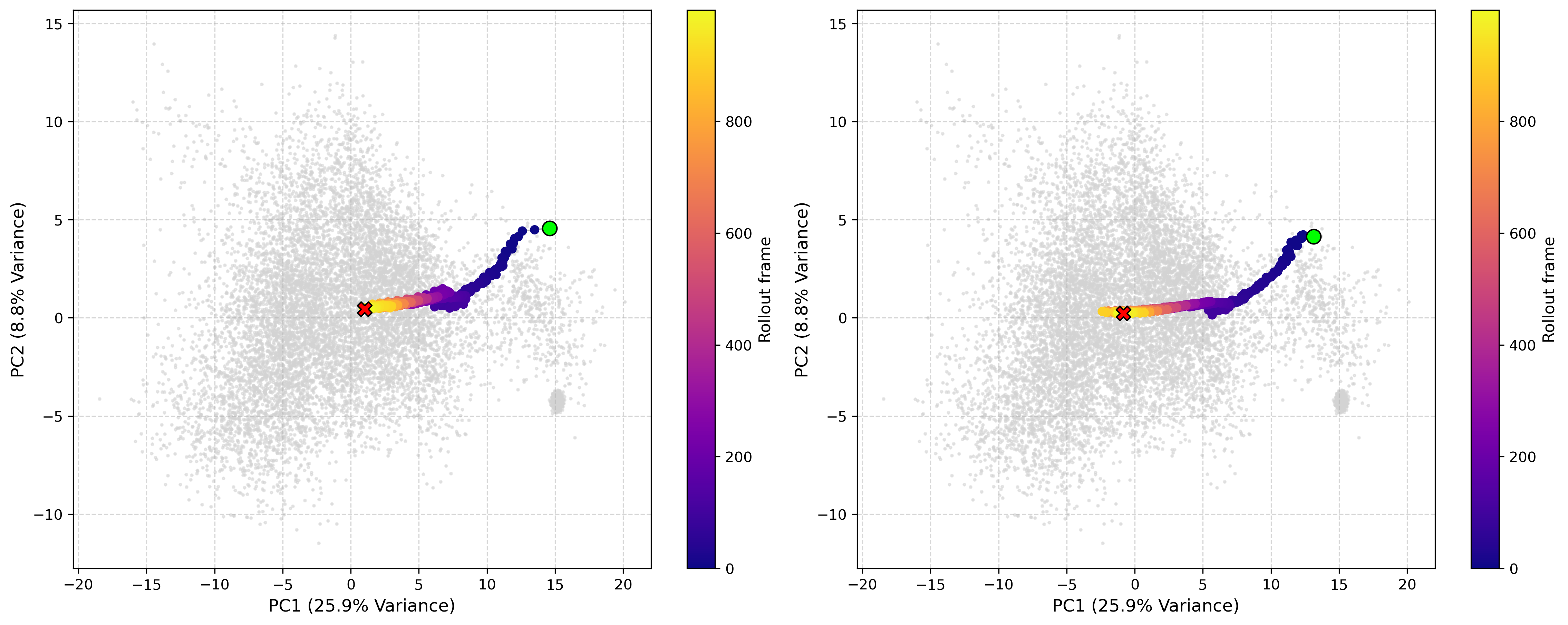}

\caption{\color{Gray} \textbf{Molecular dynamics and \texttt{iT} prediction: reproducibility using different datasets of the same size for training.} The figure shows the \texttt{iT} prediction using the series of frames ending with 1000 from simulation \#63 as a \texttt{lookback} sequence. Both \texttt{iT}s were trained with a final \texttt{lookback}/\texttt{forecast} 60/100 scheme using the curriculum learning strategy and the custom loss function (see the Methods section) and used recursively to obtain a sequence of 1000 frames. The predicted initial and final frames are shown in green and red respectively. The PCA of the simulation is shown in the gray background. \textbf{Left panel:} the \texttt{iT} prediction using the weights obtained by training with \textit{dataset1}; \textbf{Right panel:} the \texttt{iT} prediction using the weights obtained with \textit{dataset2}.}

\end{figure}
In both cases, the generated pseudo-trajectory traces an initial path along the two PCs, followed by a large number of conformations that unfold primarily along the first PC. This behavior suggests that the neural network has learned an attractor in the protein's conformational space.

The comparison of the predicted trajectories by different \texttt{iT}s, trained on datasets of different sizes, is particularly interesting. Figure 3 presents an analysis of this kind. All training parameters for the two Transformers used to predict the trajectories shown in the Figure were the same but the training dataset. The neural network used for the left panel is the one shown previously (trained on \textit{dataset1}, MAE 0.6710), while the one used for the right panel was trained on \textit{mini-dataset1} (MAE 0.6745). Although there are some differences in the trajectory predicted by the two systems, the Figure clearly shows that the overall behavior is very similar. The main difference that could be noticed is that the one derived from the smaller dataset appears to be much more "dynamic". Predicted trajectories obtained from the \texttt{iT}s that exhibited lower training errors generally show less uncertainty in their progression from the starting point to the end point and appear much more tightly clustered around a specific point at the end of the trajectory. Figure 3 also illustrates that, often, the pseudo-dynamics predicted by these \texttt{iT} models follows a trajectory that tends toward a sort of expected mean structure (compare the frames after the \texttt{lookback} sequence obtained by MD with the prediction of the two neural networks). Although a more thorough and systematic analysis is needed to clarify this behavior, it is interesting to note that there is a relationship between the pseudo-trajectory predicted by the \texttt{iT} starting from a certain trigger sequence and the trajectory that would follow the same sequence in MD. Making this aspect even more interesting is the fact that the dynamics used for these analyses were not included in the datasets used for training and evaluating the various iTs. This suggests a good generalization capability of the model, despite its low complexity given the order of magnitude of the trainable parameters. The remarkable similarity between the pseudo-trajectories obtained using two datasets of different sizes—with the smaller one requiring only very modest computation time—suggests that this type of algorithm can be effectively employed even using small MD datasets.
\begin{figure}[ht] 

\includegraphics[width=\textwidth]{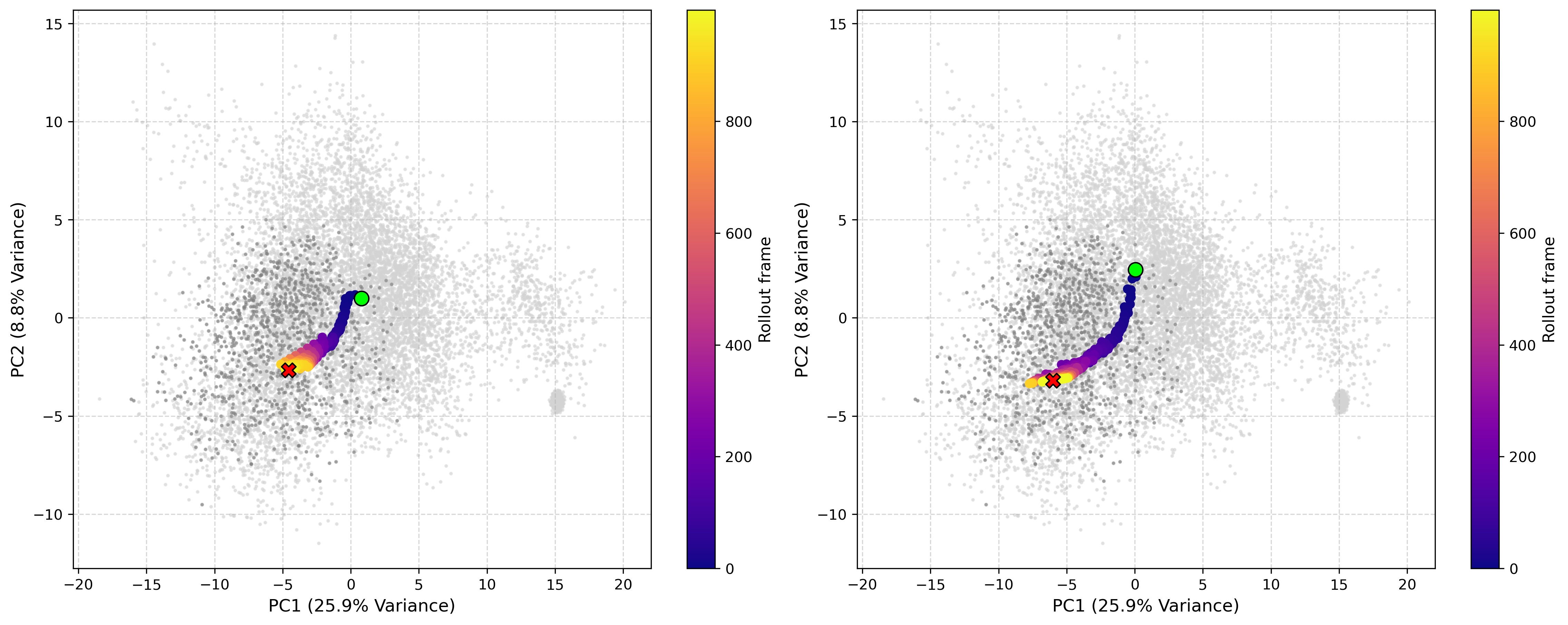}

\caption{\color{Gray} \textbf{Molecular dynamics and \texttt{iT} prediction: reproducibility using different datasets of different sizes for training.} The figure shows the \texttt{iT} prediction using the series of frames ending with 7850 from simulation \#63 as a \texttt{lookback} sequence. The PCA of the simulation is shown in the gray background. The 1000 frames following the \texttt{lookback} sequence calculated using MD are shown in dark grey. The 1000 frames of the pseudo-dynamics predicted by the \texttt{iT} are reported in color code; the predicted initial and final frames are shown in green and red respectively. \textbf{Left panel:} the \texttt{iT} prediction using the weights obtained by training with \textit{dataset1}; \textbf{Right panel:} the \texttt{iT} prediction using the weights obtained with \textit{small-dataset1}.}
\end{figure}

In this study, numerous variants of \texttt{iT} were tested. The parameters evaluated included, for example, the number of heads, the size of the networks upstream and downstream of the Transformer module, dropout, the \texttt{lookback}/\texttt{forecast} ratio, the loss function, and the training scheme. Naturally, all these parameters influence the final outcome of the training when these architectures are evaluated using a series of test simulations. It is important to note that the values obtained to assess the performance of the various architectures must be carefully interpreted within the context of the specific experiment. Generally MAE values are quite low, so these differences could be considered physically insignificant. This MAE-based performance quantification is certainly useful for distinguishing which one, among a set of Transformers trained in theoretically identical ways, offers the best performance. However, comparing the performance of different neural networks, or even the same architecture trained in completely different ways, requires additional considerations.
\begin{figure}[ht] 

\includegraphics[width=\textwidth]{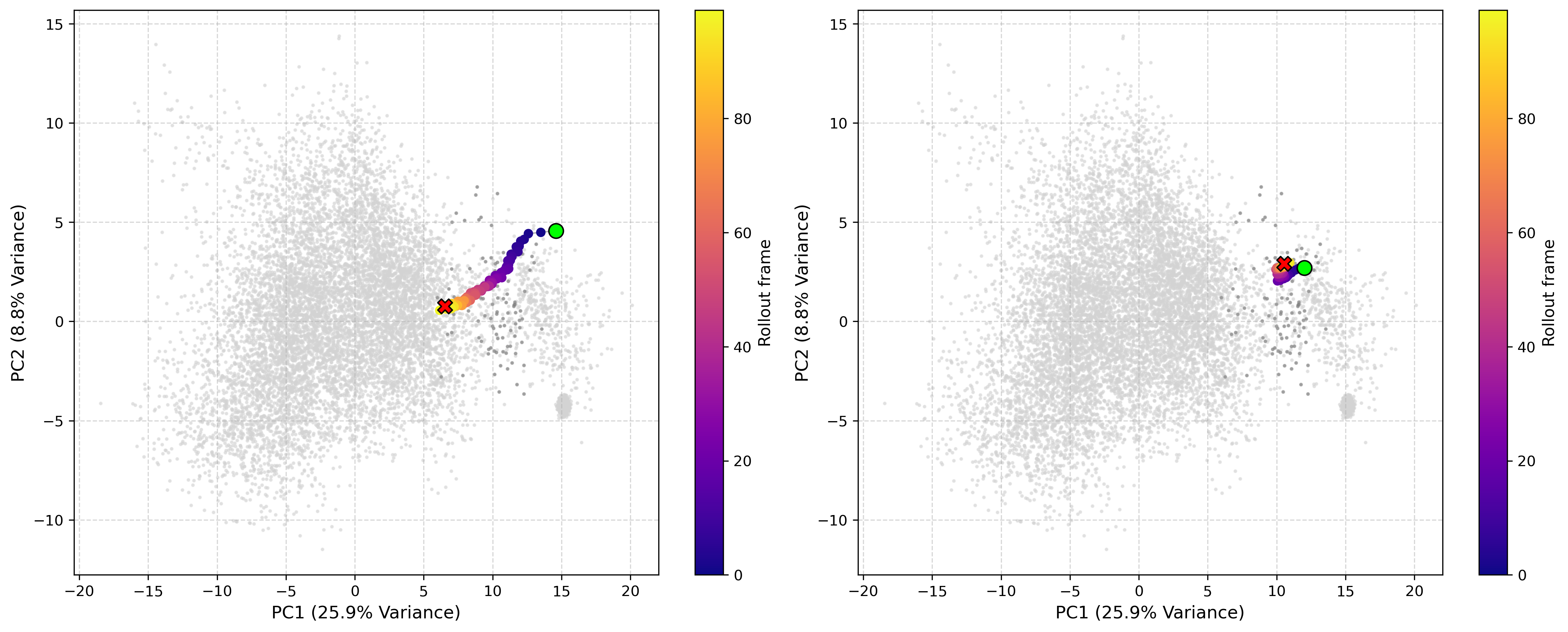}

\caption{\color{Gray} \textbf{Molecular dynamics and \texttt{iT} prediction: training effects.} The figure shows the \texttt{iT} prediction using the series of frames ending with 1000 from simulation \#63 as a \texttt{lookback} sequence. In both cases the prediction of 100 steps by the \texttt{iT} is shown. The PCA of the simulation is shown in the gray background. The 100 steps obtained by MD following the \texttt{lookback} sequence are shown in darker grey. The predicted initial and final frames are shown in green and red respectively. \textbf{Left panel:} the \texttt{iT} prediction using the weights obtained by training with \textit{dataset1} using the custom RMSF-weighted hybrid loss function and the curriculum learning strategy with a final \texttt{lookback}/\texttt{forecast} scheme 60/100; \textbf{Right panel:} the \texttt{iT} prediction using the weights obtained by training with \textit{dataset1} using a classical loss function without considering the RMSF with a fixed \texttt{lookback}/\texttt{forecast} scheme 60/10.}

\end{figure}
An example of this type of effect is shown in Figure 4: here, the same dataset was used to train two \texttt{iT} models and the skeleton of the two models was exactly the same (same number of heads, same dimension of the upstream and downstream networks), while the \texttt{lookback}/\texttt{forecast} scheme and the training method are different. In one case (left panel), the loss function employed was the one developed in this work (i.e. a combination of RMSD and RMSF) together with the curriculum learning strategy, while in the other (right panel) a standard loss function was used (essentially the same as in the previous example, but with the RMSF contribution set to zero) and the fixed \texttt{lookback}/\texttt{forecast} scheme 60/10. The MAE of the \texttt{iT} used for the left panel in the Figure was 0.6710 (see above), while that of the \texttt{iT} used for the right panel in the figure was equal to 0.5870. Obviously, in the last \texttt{iT} a smaller error is expected considering the length of the \texttt{forecast} parameter. But beside the MAE value, analyzing the Figure it can be observed that the \texttt{iT} used in the left panel allows one to obtain a prediction that not only moves significantly away from the last \texttt{lookback} frame in the plane of the first two PCs, but also recapitulates the main movements of the protein that would be obtained by performing the MD (dots in darker gray in the Figure). On the contrary, the \texttt{iT} used to obtain the right panel in the Figure seems to simply add some noise to the last structure of the \texttt{lookback} sequence, without really being able to extrapolate the expected motion of the protein into the future.

\subsection*{Protein reconstruction}
The \textit{E. coli} DHFR contains three flexible loops that form a complex dynamic network: the Met20 loop (residues 9–24), the F-G loop (residues 116–132), and the G-H loop (residues 142–150) \cite{schnell2004structure}. The Met20 loop cycles through distinct conformational states that regulate access to the active site, substrate binding, and product release. At least three distinct states can be identified crystallographically, generally designated as \textit{closed}, \textit{occluded}, and \textit{open} \cite{sawaya1997loop}. In NADPH-containing states (or states similar to those containing reagents), such as the DHFR:NADPH holoenzyme and the DHFR:NADPH:DHF Michaelis complex, the Met20 loop is in the \textit{closed} state. In this conformation, the Met20 loop blocks access to the active site, preventing solvent entry while simultaneously facilitating the correct positioning of the nicotinamide ring through a series of hydrogen bonding interactions. The \textit{occluded} conformation is observed primarily in the states occupied by the THF product, with or without the oxidized electron carrier: in this conformation, the Met20 loop protrudes into the active site, preventing access of the nicotinamide ring to the active site. Although some experimental evidence suggests that these movements may simply be due to the conformational flexibility of the \textit{E. coli} DHFR, it is generally believed these conformations are associated with the catalytic cycle of the enzyme \cite{schnell2004structure,arora2009functionally,reddish2016ligand,sawaya1997loop,liu2013temporally}. The F-G loop participates in the conformational changes of the enzyme: in particular, the \textit{closed} conformation is stabilized by the formation of hydrogen bonds between Asp122 and the Gly15 and Glu17 residues of the Met20 loop. The \textit{occluded} conformation is instead stabilized by hydrogen bonds between Asn23 of the Met20 loop and Ser148 of the G-H loop.
\begin{figure}[ht] 
\includegraphics[width=\textwidth]{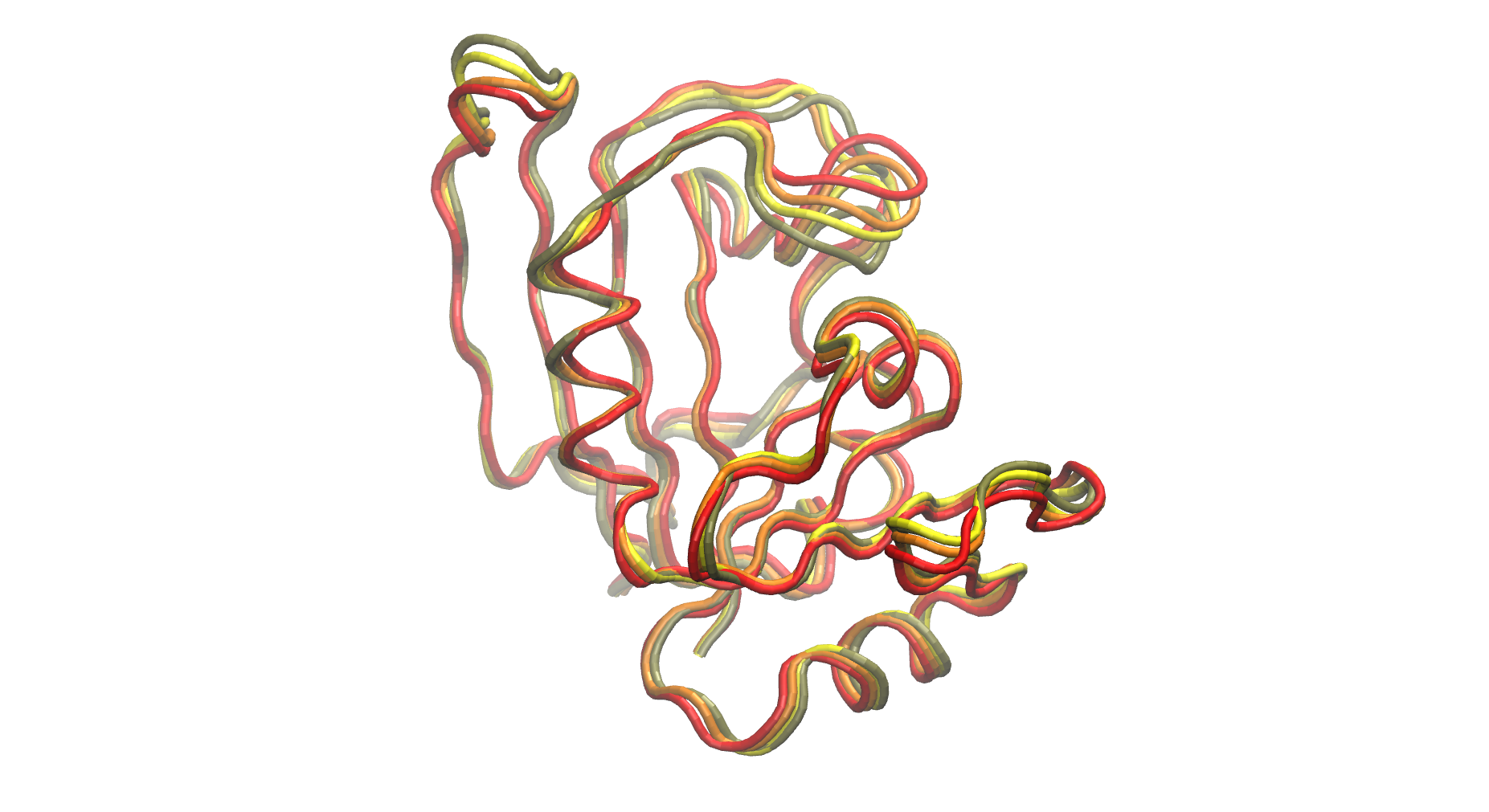}
\caption{\color{Gray} \textbf{Structures predicted by \texttt{iT}.} The figure shows the C$\alpha$ atom skeletons predicted using an \texttt{iT} in generative mode. The structures were sampled from the predicted sequence using the weights of an \texttt{iT} trained with \textit{dataset1} using the custom loss function and the curriculum learning strategy. These are the same structures projected into the PCA shown in the left panel of Figure 1 obtained using the series of frames ending with 1000 from simulation \#63 as a \texttt{lookback} sequence. The first structure of the predicted sequence is shown in red, the structure number 10 in the sequence is shown in orange, the structure number 100 in yellow, and the final structure of the predicted sequence is shown in tangerine.}
\end{figure}
Since the dynamics of various regions of the \textit{E. coli} DHFR have been studied in detail, it is interesting to see whether the DHFR dynamics predicted by the \texttt{iT} are able to replicate the movements of the various loops of the protein. As shown in Figure 5, the dynamics obtained using an \texttt{iT} trained as described in this work shows movements that exactly mirror the known motility of the protein. The neural network used is able to perfectly predict the movement of the Met20 loop, which passes through a series of conformations that are very similar to the conformations described both by X-ray crystallography and computationally. The regions corresponding to the G-H and F-G loops also show a series of movements comparable to those previously described in the literature. The Figure also shows that small fluctuations in the region around residue 65, a key area of the protein close to the NADPH binding site, similar to what can be observed in MD.
This analysis demonstrates that \texttt{iT} is able to accurately predict the main motions of the protein, in accordance with those described in a large literature on this model protein.

\subsection*{Did \texttt{iT} learn protein thermodynamics?}
As shown above, the \texttt{iT} is able to predict an ordered sequence of DHFR structures that arrange themselves in a kind of pseudo-dynamics, mirroring those experimentally observed. This ordered sequence of structures generally ends up in highly populated regions of the conformational landscape. The analyses reported above refer to single sequences and cannot be used to make thermodynamic inferences. But the GenMD requires obtaining not only admissible protein conformations, avoiding distortions and alterations that lead to the prediction of a non-physically coherent structure: true GenMD systems should be able to determine the statistical weight of the admissible conformations.
\begin{figure}[ht] 
\centering
\includegraphics[width=0.75\textwidth]{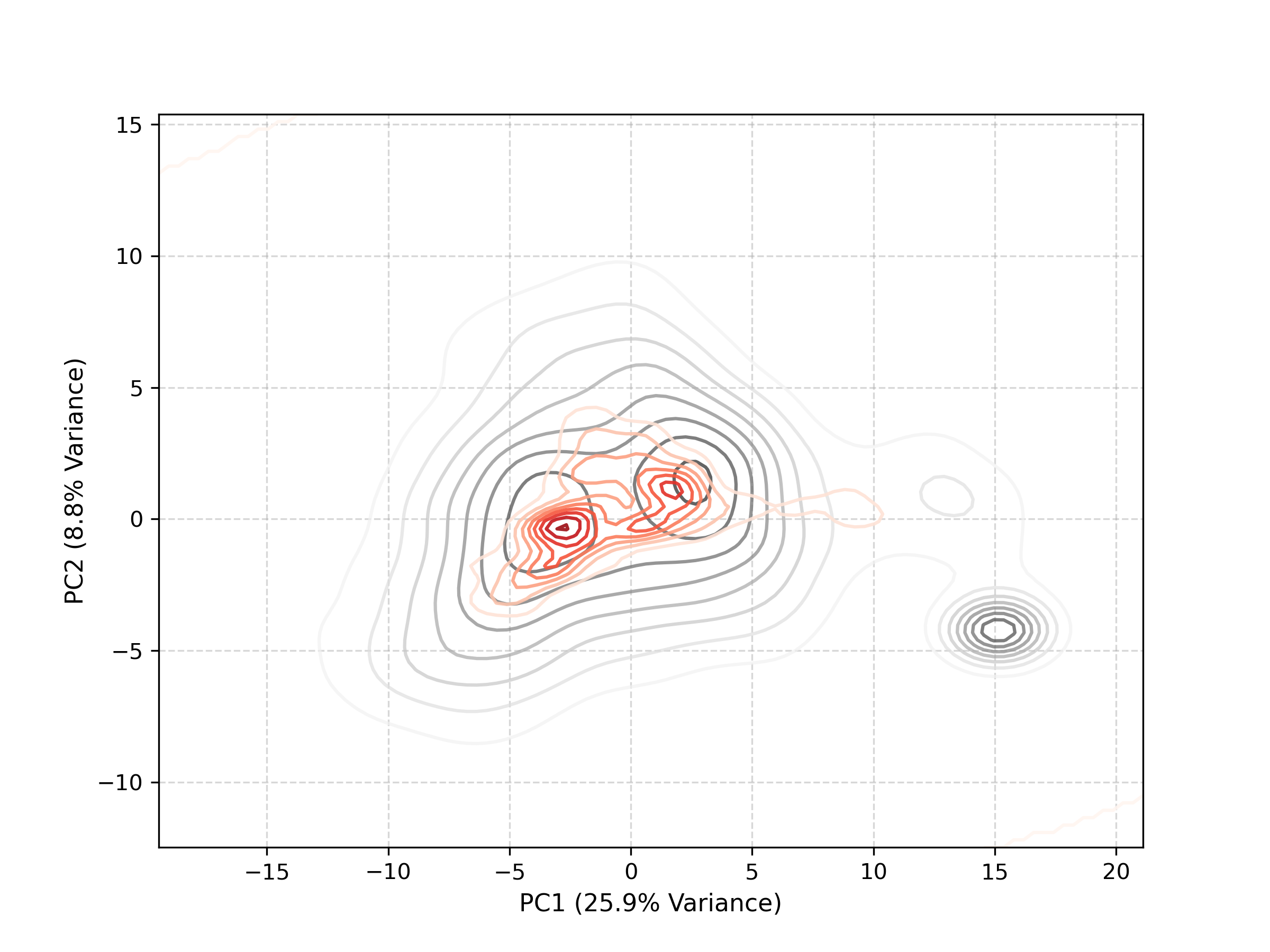}
\caption{\color{Gray} \textbf{Probability of DHFR conformations.} The probability of the protein conformations obtained by MD is reported as a grayscale, the probability of the protein conformations predicted by \texttt{iT} is reported as a redscale. All the conformations used were projected onto the plane identified by the first two PCs obtained by performing PCA on the conformations derived from MD. The simulation used both for the PCA calculation and to obtain the \texttt{lookback} sequences provided as input to the neural network shown in the figure is simulation \#63, which was not part of any training dataset. The MD shows two high-probability regions symmetrically distributed around the origin of the axes (the high-probability region of the MD-derived conformations observed in the lower right quadrant is the one in which the conformations close to the starting crystallographic structure obtained in the equilibration phase are found; these initial conformations are excluded from the MD-derived conformations extracted for the construction of the training datasets). The two probability peaks of the production run coincide almost perfectly with the probability peaks of the conformations predicted by the neural network. The weights used for the prediction are the same as the \texttt{iT} used to obtain Figure 1, left panel.}
\end{figure}
It would be interesting to understand whether \texttt{iT}, in addition to having learned how to reproduce an optimized sequence of admissible structures, is able to obtain a measure of the probability of the protein conformations themselves, because this would be the basis for considering this neural network architecture as a possible candidate for effective GenMD.  As mentioned above, the terminal conformations of a predicted pseudo-dynamics obtained recursively using the same \texttt{iT} prediction as a trigger are in fact oscillations around an average conformation. This might suggest that the \texttt{iT} is able to predict a statistically probable terminal conformation and that once this conformation is reached, the \texttt{iT} simply adds background noise. However, this noise is not the thermal noise experienced by the real protein (and simulated by the MD), and hence the impossibility of escaping from a sort of potential well. While this analysis of individual trajectories suggests that the neural network may indeed have somehow learned the statistical weight of the admissible conformations, performing the prediction recursively (i.e. repeatedly using sequences of protein conformations predicted by the same \texttt{iT} as \texttt{lookback} sequences) exposes the risk of considering as statistically highly probable protein configurations that are actually repeatedly predicted simply because the \texttt{iT} ended up in an attractor (in the mathematical sense of the term).

To evaluate the ability of \texttt{iT} to encode a statistical weight to the protein conformations it predicts, a detailed analysis of the terminal conformations produced by the neural network was performed. All the \texttt{lookback} sequences obtainable with the sliding window technique (i.e., the set of \texttt{lookback} sequences that end at the various time steps $\{t, t + 1, t + 2, \dots, t + T\}$, where $T$ is the total number of time steps included in the analysis) were extracted from a simulation not included in the training dataset. All these sequences were used as input for the \texttt{iT}, and the last conformation obtained as output was inserted into a dataset of predicted structures. The PCA of the MD run used to obtain the \texttt{lookback} sequences was calculated, and the conformations obtained via MD were projected onto the plane defined by the first two PCs. The conformations calculated by the \texttt{iT} and grouped into the set described above were projected onto this same plane. The probability distribution on this plane of the conformations obtained by MD and the probability distribution of the conformations obtained as output from the \texttt{iT} were then calculated. Figure 6 shows one of these analyses: as can be seen from the Figure, in both cases a probability distribution is obtained with two very sharp maxima within a broader probability maximum. Particularly interesting is the overlap of the two distributions: the maximum probability peaks (corresponding to the protein configurations at their energetic minimum) are practically identical. These results strongly suggest that the \texttt{iT} has learned not only the admissible conformations, but has also somehow learned the thermodynamics of the system. Furthermore, the protein conformations contained in this \texttt{iT} output set, if ordered in the same sequence as the MD conformations, constitute a pseudo-dynamics that very closely resembles true molecular dynamics (not shown).
\section*{Discussion}
The development of computational protocols based on AI algorithms is an extremely active field of research. The promise of obtaining AI models capable of predicting the possible conformations of a protein would be the culmination of the revolution initiated by the introduction of algorithms capable of generating a protein's three-dimensional structure from its sequence. 
This paper illustrates a simple neural network model capable of easily learning the coarse-grained dynamics of a specific protein. The neural network employed belongs to the Transformer family \cite{vaswani2017attention}, but it has the unique characteristic of using, as tokens, time-ordered sequences \cite{liu2023itransformer} of the various degrees of freedom $V$ that describe the motion of proteins in space. The representation of protein motion thus consists simply of the Cartesian coordinates of the atoms of interest—specifically, the C$\alpha$ atoms. The attention matrix was constructed to have a dimension of exactly $V \times V$ for each attention head and each batch. This is the distinctive and innovative feature of the \texttt{iT}: in classical Transformers, tokens represent time steps, and the attention matrix is sized according to the length of the time sequence (i.e., $time \times time$). In the \texttt{iT}, because the axes have been swapped via a permutation operation, each variable (each atom/coordinate) becomes an independent token. Consequently, the attention matrix $A$ relates every variable to all others, resulting in a square map that captures the correlations and spatial cooperativity among all the atomic coordinates of the protein. 

Considering the data reported in the Results section, an interesting picture of the \texttt{iT}'s behavior emerges. The appearance of the predicted conformations indicates that the model has not simply memorized the training trajectory, but has extracted the essential geometry of the collective motion of the C$\alpha$ atoms, a sort of distillation of the manifold intrinsic to the protein's dynamics. Another extremely important aspect is the matching of the probability maxima: observing the contour plot of the probability distribution of the structures obtained by MD and predicted by the \texttt{iT}, it is clear that the probability maxima coincide. The \texttt{iT} is not guessing or generating random coordinates (which would be associated with structures lacking physical coherence), but has learned the protein's thermodynamics and the topology of the dominant conformers, recognizing with great precision the main energy minima where the protein spends most of its time in the real simulation. Another significant aspect is that the model does not exhibit uncontrolled numerical drift, remaining confined to the physically plausible regions explored by the MD. This result can be attributed to the hybrid loss function used and the Z-score normalization. Indeed, in addition to the overall architecture of the \texttt{iT}, and the consequent radical revision of the token concept in protein dyanamics, it is worth highlighting the crucial role of the hybrid loss ($\mathcal{L}_{\text{MSE}} + \lambda \mathcal{L}_{\text{RMSF}}$) introduced in this work: many generative models fail in the long run because they lose variance (thermal fluctuations), flattening out on a static mean. The explicit penalty on the RMSF forced the network to preserve the amplitude of the fluctuations, preserving the (thermo)dynamics of the system. 

Further supporting the idea that \texttt{iT} has truly learned the protein's dynamics is the fact that the tokens used as input and imposed as output are tensors whose members can have any numeric value. Unlike classical Transformers, a finite number of tokens was not imposed, nor were artificial words created to compose a pseudo-language of the protein. Although the number of input and output tokens is, in effect, infinite, the \texttt{iT} produces configurations that are physically plausible and with a frequency that reflects the expected thermodynamic weight, without exploding (or imploding) numerically: the most plausible explanation is, again, that it has effectively learned the dynamics of the \textit{E. coli} DHFR.

A consideration should also be made regarding computational efficiency: being able to replicate energy minima and basins of attraction on a completely unprecedented trajectory (held-out test set), using an extremely lightweight model (less than a million parameters), is certainly a significant achievement. Currently, the literature on the application of AI algorithms to protein dynamics moves in two main directions: machine learning force fields or GenMD models based on flows or large-scale diffusion. These models require tens or hundreds of millions of parameters, massive GPU clusters, and mammoth datasets. Large foundational models or models based on diffusion/flow matching are able to sample the global thermodynamic equilibrium in a significant way, but often at the cost of daunting computational complexity and near-zero interpretability (they are impenetrable "black boxes"). Furthermore, they are prone to "over-parameterization", which makes them prone to accumulate numerical drift or collapse into non-physical structures (out-of-distribution drift) when unleashed in prolonged rollouts, unless rigorous $SE(3)$ geometric invariance constraints are imposed. The inverted Transformer approach (which treats spatial coordinates as tokens and the time axis as features) together with hybrid loss and curriculum learning  is disarmingly computationally parsimonious, managing to capture global correlations between residues without the need for an explicit and (computationally very heavy) 3D geometric architecture.

Although only the \textit{E. coli} DHFR apoprotein has been discussed in detail in this communication, the same computational protocol has been applied to other protein systems that will be reported and discussed elsewhere (among them the \textit{E. coli} DHFR with different ligands, the human DHFR and the lysozyme of the bacteriophage T4). The \texttt{iT} algorithm is extremely robust, meaning that, beyond the inevitable small differences in behavior, the predictions of all the \texttt{iT} algorithms shown are extremely similar to each other (in fact, from the point of view of the biochemical significance of the obtained data, essentially identical). The results presented here certainly do not represent the definitive answer to the search for the final GenMD algorithm: the algorithm was applied to the MD data of a single protein, and moreover, using a coarse-grained representation of that protein. Despite this, the algorithm is interesting for its simplicity and efficiency, which allows the interested researcher to expand the number of structures available at a very limited computational cost. All this can be achieved even using a very small dataset of MD structures, as demonstrated above. Finally, and perhaps most interestingly, the generalization capability of \texttt{iT} makes it an excellent candidate for a truly efficient GenMD system. The next step will be to insert \texttt{iT} modules into larger models encoding the protein sequence. Such \texttt{iT} models, capable of being trained on multiple different proteins, could form the basic architecture of future systems that are really able to perform the GenMD.


\section*{Code availability}
The \texttt{iTransformer} architecture and the recursive dynamics forecasting pipeline were implemented in Python 3.11.14 using the Keras/JAX framework. Custom scripts for data processing and structural analysis were developed with the support of Gemini (Google) and will be available at https://github.com/llpalese/MDiT.



\section*{Acknowledgments}
The author thanks the EuroHPC Joint Undertaking for awarding this project access to the EuroHPC supercomputer LEONARDO, hosted by CINECA (Italy) and the LEONARDO consortium through an EuroHPC Development Access call (Project EUHPC\_D27\_087). The author thanks Google Gemini, which provided support in software development, code refactoring, and technical debugging during the preparation of this work.


\section*{Supporting Information}

\setcounter{figure}{0}
\renewcommand{\thefigure}{S\arabic{figure}}

Below is the composition of the various datasets mentioned in the main text. A total of 66 simulations were used, distributed as follows:
\begin{itemize}
    \item \textit{dataset1}
        \begin{itemize}
            \item training data - simulations 1-25
            \item validation data - simulations 26-29
            \item test data – simulation 30
        \end{itemize}
    \item \textit{dataset2}
        \begin{itemize}
            \item training data - simulations 31-55
            \item validation data - simulations 56-99
            \item test data – simulation 60
        \end{itemize}
    \item \textit{mini-dataset1}
        \begin{itemize}
            \item training data - simulations 1-5
            \item validation data - simulations 26-27
            \item test data – simulation 30
        \end{itemize}
    \item Simulations not included in the training-validation-test datasets
        \begin{itemize}
            \item simulations 61-66
        \end{itemize}
    
\end{itemize}

For all these simulations, RMSD, RMSF, DCCD, and PCA were calculated. Examples of these analyses are shown in the figures below for the two simulations outside the training sets mentioned in the text.

\begin{figure}[ht] 
\centering
\includegraphics[width=\textwidth]{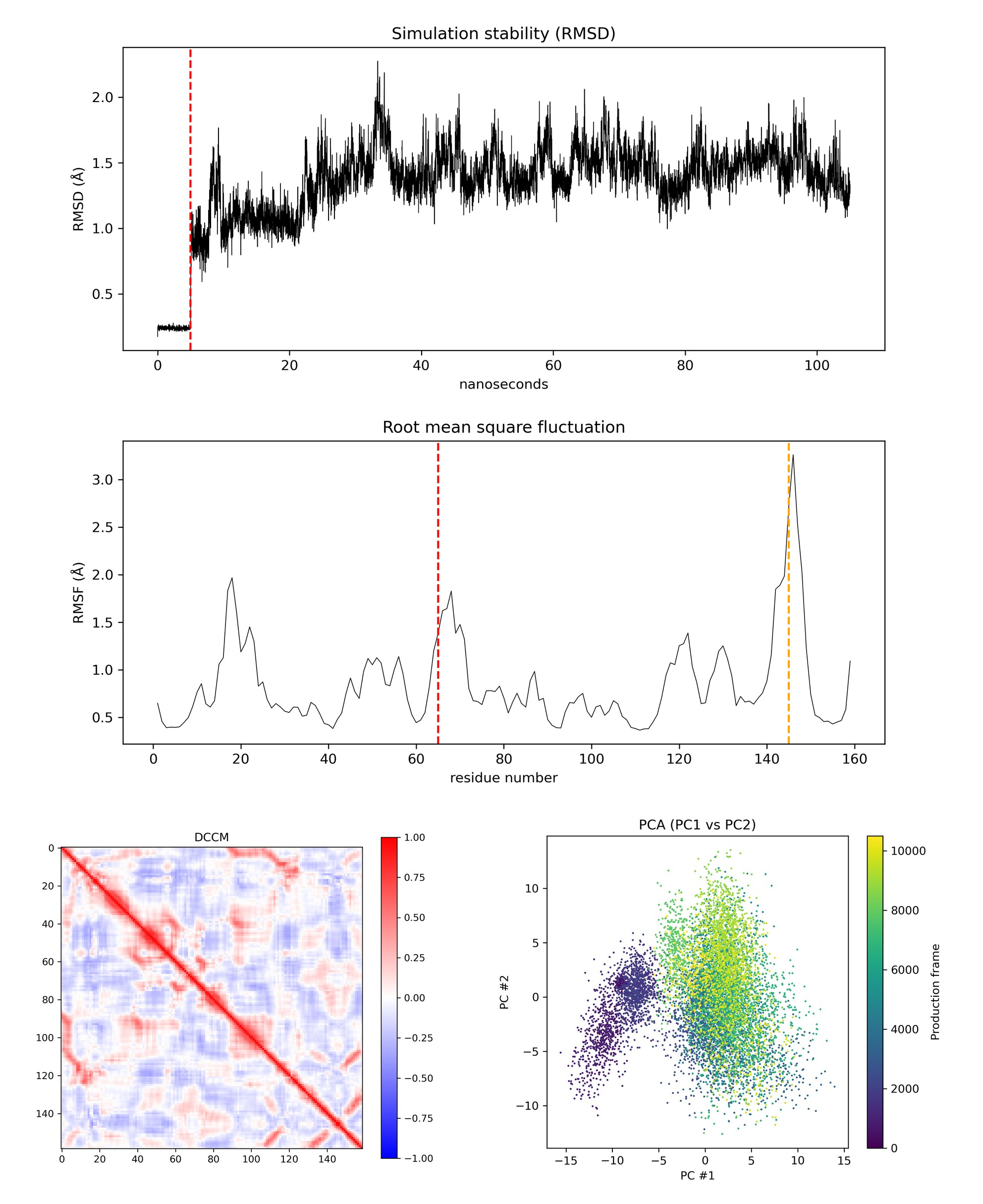}
\caption{\color{Gray} \textbf{Simulation \#61 of the \textit{E. coli} DHFR dataset.} The top panel reports the RMSD values (the red vertical line represents the cutoff of the simulation step selected for training); the middle panel reports the RMSF values (the red and orange lines represent the positions of residues 65 and 145); the bottom left panel reports the Dynamic Cross-Correlation Matrix; the bottom right panel reports the Principal Component Analysis.}
\end{figure}

\begin{figure}[ht] 
\centering
\includegraphics[width=\textwidth]{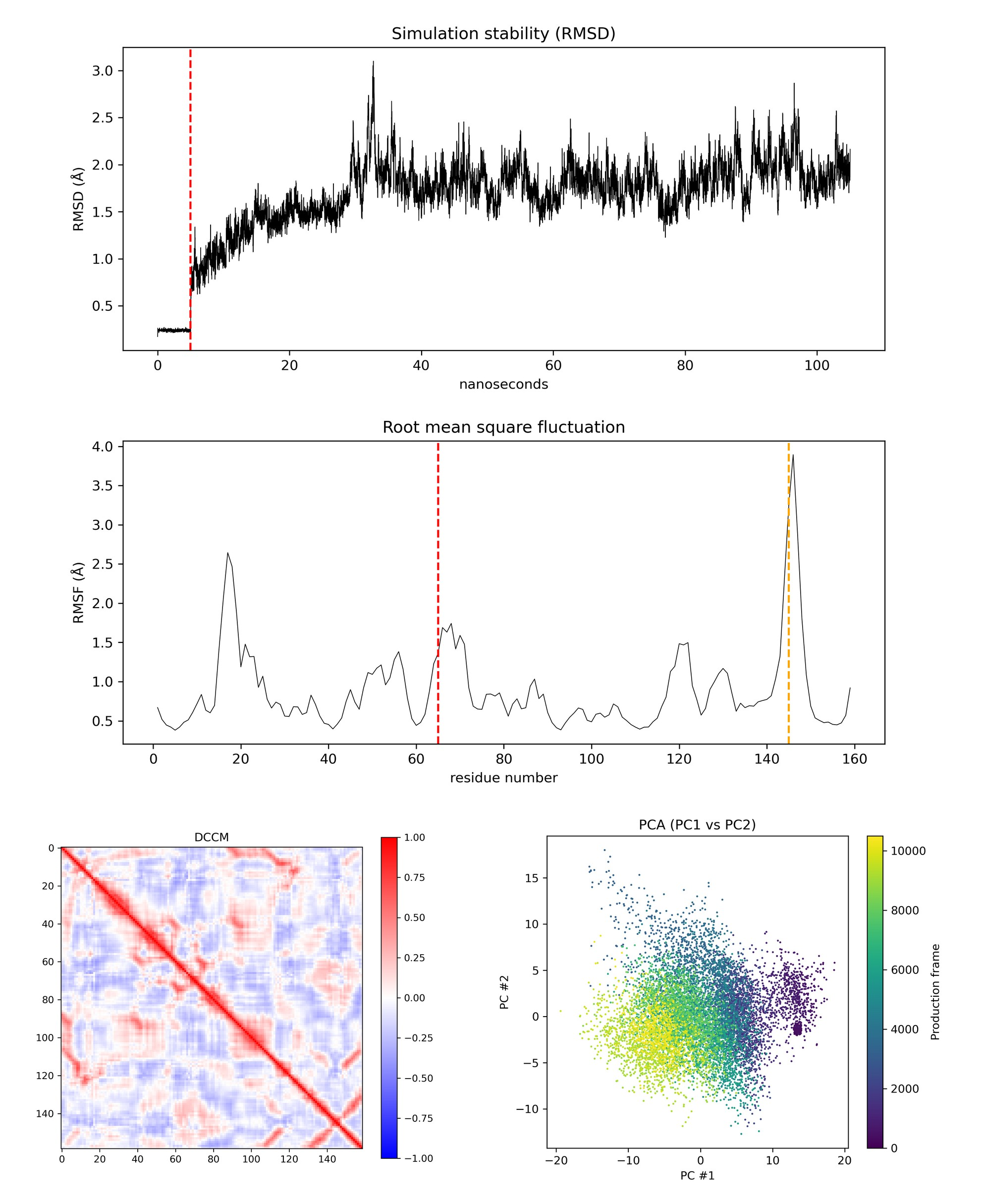}
\caption{\color{Gray} \textbf{Simulation \#63 of the \textit{E. coli} DHFR dataset.} The top panel reports the RMSD values (the red vertical line represents the cutoff of the simulation step selected for training); the middle panel reports the RMSF values (the red and orange lines represent the positions of residues 65 and 145); the bottom left panel reports the Dynamic Cross-Correlation Matrix; the bottom right panel reports the Principal Component Analysis.}
\end{figure}


\clearpage

\nolinenumbers

\bibliography{library}

\bibliographystyle{abbrv}

\end{document}